\documentclass[aps, reprint,pra,nofootinbib,longbibliography]{revtex4-1}
\usepackage{dcolumn}
\usepackage[utf8]{inputenc}
\usepackage{amsmath}
\usepackage{amsfonts}
\usepackage{amssymb}
\usepackage{graphicx}
\usepackage{hyperref}
\usepackage{xcolor}

\newcommand{\mb}{\mathbf}
\newcommand{\bs}{\boldsymbol}

\begin{document}

\title{Electrodynamic Aharonov-Bohm effect}

\author{Pablo L. Saldanha}\email{saldanha@fisica.ufmg.br}
\affiliation{Departamento de F\'isica, Universidade Federal de Minas Gerais, Belo Horizonte, MG 31270-901, Brazil}

\date{\today}

\begin{abstract}
We propose an electrodynamic Aharonov-Bohm (AB) scheme where a nonzero AB phase difference appears even if the interferometer paths do not enclose a magnetic flux and are subjected to negligible scalar potential differences during the propagation of the quantum charged particle. In the proposal, the current in a solenoid outside the interferometer varies in time while the quantum particle is in a superposition state inside two Faraday cages, such that it is always subjected to negligible electromagnetic fields. At first glance, this result could challenge the topological nature of the AB effect. However, by considering the topology of the electromagnetic field configuration and the possible particle trajectories in spacetime, we demonstrate the topological nature of this situation. 
\end{abstract}


\maketitle

In the magnetic Aharonov-Bohm (AB) effect, the interference pattern of quantum charged particles depends on the magnetic flux enclosed by the interferometer paths even if the particles only propagate in regions with null electromagnetic fields \cite{ehrenberg49,aharonov59}. This effect was experimentally observed with many different systems \cite{chambers60,tonomura86,oudenaarden98,bachtold99,ji03,peng10,becker19,nakamura19,deprez21,ronen21} and has a profound fundamental significance in Physics, since it contradicts the notion that an electric charge is only affected by the local electromagnetic fields. 

The magnetic AB effect has a topological nature \cite{peshkin95,cohen19}, as discussed in textbooks \cite{ballentine,bransden,gottfried}, since it is not possible to deform one of the interferometer paths into the other, while keeping its initial and final positions, without crossing a region with nonzero magnetic fields. The region with a null field is not simply connected. Also, the AB phase difference depends only on the magnetic flux enclosed by the two interferometer paths, not on the specific paths geometry. 

But here we predict nonzero AB phase differences when the electric current in a solenoid outside the interferometer varies in time while the quantum particle is in a superposition state inside two Faraday cages, with the particle paths enclosing no magnetic flux and being subjected to a negligible scalar potential difference. This result could challenge the topological nature of the AB effect. However, by considering the topology of the spacetime region with a null electromagnetic field and the possible particles trajectories also in spacetime, not only in space, we demonstrate the topological nature of this situation. 

\begin{figure}
  \centering
    \includegraphics[width=8.5cm]{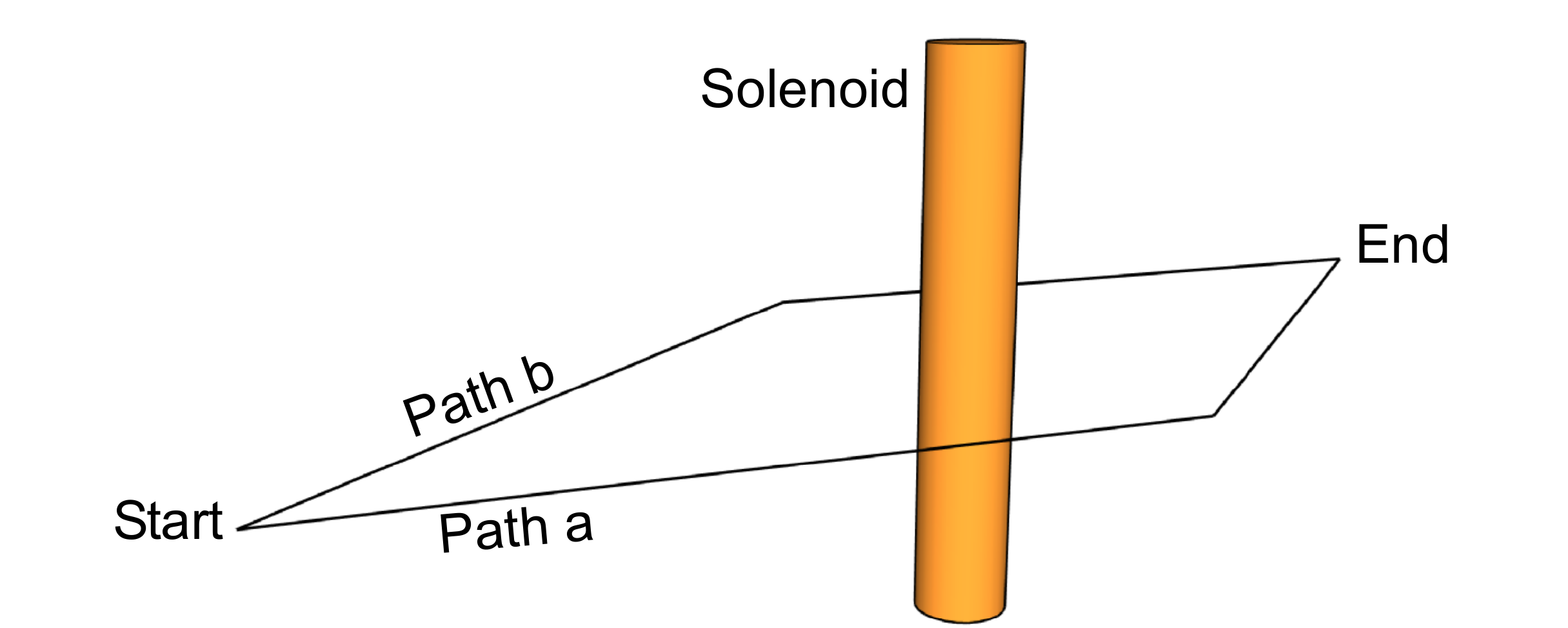}
  \caption{{Magnetic AB effect.} A quantum charged particle has two possible paths in an interferometer that encloses an infinite solenoid. The AB phase difference between the paths depends on the solenoid magnetic flux, with the particle only propagating in regions with negligible electromagnetic fields.}
\end{figure}

\begin{figure*}
  \centering
	    \includegraphics[width=18cm]{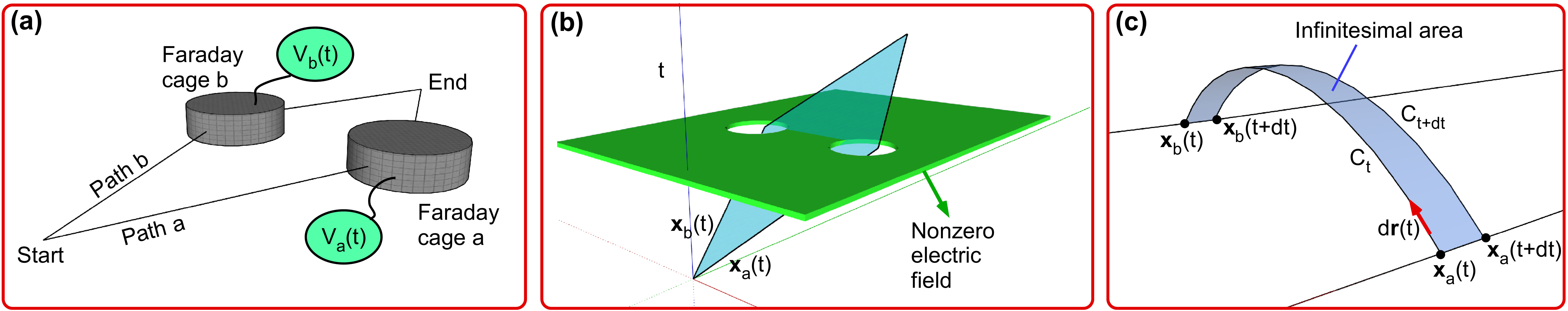}
  \caption{{Electric AB effect.} {(a)} A quantum charged particle has two possible paths in an interferometer, each one with a Faraday cage. When the particle is in a superposition state inside both cages, their potentials $V_a$ and $V_b$ vary and come back to zero before each component of the wave function leaves the corresponding cage. The AB phase difference depends on the potential difference between the paths during the particle propagation. {(b)} Possible spatiotemporal trajectories $\mb{x}_a(t)$ and $\mb{x}_b(t)$ of the particle in the interferometer, in black. We are only showing the spatial $xy$ plane in the figure, assuming that the particle trajectories are in this plane. The time coordinate is represented in the vertical direction. The spacetime region with a nonzero electric field is represented in green. One of the possible spacetime surfaces whose boundaries are the trajectories $\mb{x}_a(t)$ and $\mb{x}_b(t)$ is represented in blue. {(c)} Construction of the spacetime surface $\mathcal{S}$, whose boundaries are the trajectories $\mb{x}_a(t)$ and $\mb{x}_b(t)$. This figure is in 3D spatial coordinates. For each time $t$, we define a continuous curve $C_t$ in space  that goes from the position $\mb{x}_a(t)$ to the position $\mb{x}_b(t)$ through infinitesimal displacements $d\mb{r}(t)$. The spacetime surface $\mathcal{S}$ is formed by connecting consecutive curves, as illustrated here between times $t$ and $t+dt$. The spatial component of the formed infinitesimal spacetime surface is also defined in the figure.}
\end{figure*}

Consider that we have an interferometer for a quantum non-relativistic charged particle, which may propagate through two paths $a$ and $b$. If the particle is subjected to nonzero vector and scalar potentials $\mb{A}$ and $V$, but always propagates in regions with null electromagnetic fields, it suffers no electromagnetic force. However, the term $H_I = qV-(q/m)\mb{p}\cdot\mb{A}$
in the system Hamiltonian, where $q$, $m$, and $\mb{p}$ are the particle charge, mass, and momentum respectively, couples the quantum particle charge to the electromagnetic potentials. This coupling may result in a nonzero phase difference between the two particle paths in the interferometer, in the so-called AB effect \cite{ehrenberg49,aharonov59}. Consider that the particle wave packets are reasonably well localized along positions $\mb{x}_i(t)$ in each path during their propagation, with $i=\{a,b\}$. In this case, the AB phase accumulated in each path can be written as
\begin{equation}\label{AB_i}
	\phi_{i}=-\int_{t_0}^t \frac{H_I}{\hbar} dt'=-\frac{q}{\hbar}\int_{t_0}^t Vdt' +\frac{q}{\hbar}\int_{\mb{x}_0}^{\mb{x}_i}\mb{A}\cdot d\mb{x},
\end{equation}
where $\mb{p}/m$ was substituted by the average wave packet velocity $d\mb{x}_i/dt$, we have $\mb{x}_a(t_0)=\mb{x}_b(t_0)=\mb{x}_0$ before the wave function splitting in the interferometer, and the spatial integral is performed through the particle path. In all situations to be treated here, we assume that the interferometer lengths are small and the timescales are large, such that the fields propagation times within the interferometer can be disregarded.

Fig. 1 illustrates a scheme that demonstrates the magnetic AB effect \cite{ehrenberg49,aharonov59}. An infinite solenoid with magnetic flux $\Phi_0$ produces null electromagnetic fields in the particle paths. However, the particle is subjected to a nonzero vector potential. Using Eq. (\ref{AB_i}), Stokes' theorem, and the relation $\mb{B}=\bs{\nabla}\times\mb{A}$ for the solenoid magnetic field, the AB phase difference between the paths is given by 
\begin{equation}\label{AB_B}
	\phi_{AB}^{(1)}=\frac{q}{\hbar}\oint\mb{A}\cdot d\mb{x}=\frac{q}{\hbar}\int_S \mb{B}\cdot d\mb{a}=\frac{q\Phi_0}{\hbar},
\end{equation}
with the surface $S$ of the area integral having the particle paths as its boundary. The result is the same for a very long solenoid, with the particle propagating in regions with negligible fields. Experiments were done with a toroidal magnetic covered by a superconductor \cite{tonomura86}, guaranteeing that the particles only propagate in regions with negligible fields, confirming the predicted AB phase difference. 


\begin{figure*}
  \centering
    \includegraphics[width=18cm]{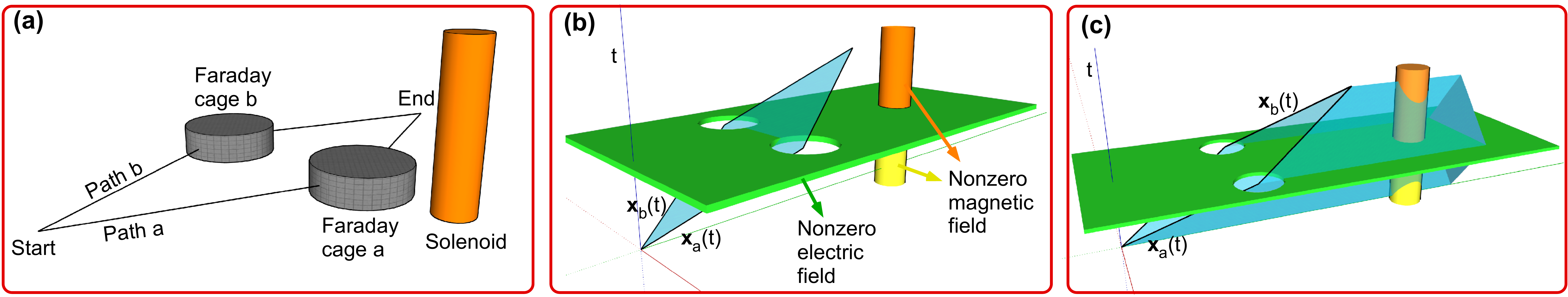}
  \caption{{Proposed electrodynamic AB scheme.} {(a)} A charged particle propagates through the interferometer, with two possible paths. While it is in a superposition state inside both Faraday cages, the solenoid current linearly changes. The AB phase difference between the paths in general is nonzero. {(b)} Possible spatiotemporal trajectories $\mb{x}_a(t)$ and $\mb{x}_b(t)$ of the particle in the interferometer, with the same representation as in Fig. 2(b). The spacetime region with a nonzero electric field is represented in green and the regions with nonzero magnetic field are in yellow and orange. One of the possible spacetime surfaces whose boundaries are the trajectories $\mb{x}_a(t)$ and $\mb{x}_b(t)$ is represented in blue. {(c)} The same as in (b), but with a different spacetime surface formed by a different deformation of one trajectory into the other.}
\end{figure*}

Fig. 2(a) illustrates the electric AB effect \cite{aharonov59}. A quantum particle is sent to the interferometer. The scalar potentials $V_a$ and $V_b$ of the Faraday cages are always zero while the quantum particle is outside them. But while the quantum particle is in a superposition state  inside both cages, the potentials $V_a$ and $V_b$ vary and come back to zero before the particle wave function leaves them. So, the quantum particle is always subjected to null electromagnetic fields. Using Eq. (\ref{AB_i}), we conclude that the AB phase difference between the paths is given by  
\begin{equation}\label{AB-E1}
	\phi_{AB}^{(2)}=-\frac{q}{\hbar}\int (V_a-V_b)dt.
\end{equation}

This electric AB effect is also a topological effect \cite{singleton13,sanders14}. To see that, we must consider the topology of the electric field configuration and the particle possible trajectories in spacetime, not only in space. Fig. 2(b) shows the possible spatiotemporal trajectories of the particle in the interferometer and the relevant behavior of the electric field in spacetime. The spacetime region with a nonzero electric field is represented in green, such that the region with a null field is not simply-connected. By deforming one of the spatiotemporal trajectories into the other in Fig. 2(b), while keeping the initial and final points of these trajectories, we define a spacetime surface $\mathcal{S}$, for instance the blue one in this figure. The general construction of these spacetime surfaces is described in Fig. 2(c). We can write the potential difference between the Faraday cages $a$ and $b$ at time $t$ as $	V_a(t)-V_b(t)=\int_{\mb{x}_a}^{\mb{x}_b}\mb{E}(\mb{r},t)\cdot d\mb{r}(t)$, $\mb{E}(\mb{r},t)$ being the electric field generated by the electric charges in the cages. Note that this integral is the same irrespective of the curve that goes from any point $\mb{x}_a$ inside Faraday cage $a$ to any point $\mb{x}_b$ inside Faraday cage $b$. So, according to Eq. (\ref{AB-E1}) the AB phase difference is given by the following integral at the spacetime surface defined by the cited path deformation \cite{singleton13}:
\begin{equation}\label{AB_E}
	\phi_{AB}^{(2)}=-\frac{q}{\hbar}\int \int_\mathcal{S} dt\, d\mb{r}(t)\cdot \mb{E}(\mb{r},t).
\end{equation}
This AB phase difference does not depend on the specific  geometry of the possible particle trajectories in spacetime nor on the specific spacetime surface $\mathcal{S}$ that is formed with the cited path deformation, as long as the spacetime surface has the possible particle trajectories as its boundary. It depends only on the topology of the spacetime region with a null electric field and on general features the particle possible trajectories in spacetime. If one possible trajectory crosses one of the ``holes'' in spacetime with no electromagnetic field  in Fig. 2(b) and the other possible trajectory crosses the other ``hole'', the above AB phase difference appears. For this reason, the electric AB effect is also a topological effect.

Now let us introduce the situation depicted in Fig. 3(a), which corresponds to an electrodynamic AB effect. The electric current in the infinite solenoid outside the interferometer linearly changes while the quantum particle is in a superposition state inside the Faraday cages, generating an electric field in the interferometer region. Charges are induced in the cages to cancel the total electric field inside them, such that the quantum particle is always subjected to negligible electromagnetic fields. Consider that the wave packets in each path are close to the positions $\mb{R}_a$ and $\mb{R}_b$ inside the Faraday cages while the solenoid current changes, not moving much. The total AB phase difference  can then be computed from 3 terms, according to Eq. (\ref{AB_i}). Let us consider the Lorenz gauge to compute it. The first term results from the interaction of the quantum particle with the initial vector potential $\mb{A}_i$ while its wave function propagates from the start to the superposed positions $\mb{R}_a$ and $\mb{R}_b$, being given by $(q/\hbar)[\int_{\mb{R}_b}^{\mb{R}_a} \mb{A}_i\cdot d\mb{x}]_\mathrm{st.}$, with the integral performed from the path that goes from $\mb{R}_b$ to $\mb{R}_a$ passing through the start position in Fig. 3(a). A second term results from the interaction of the particle with the scalar potential resultant from the electric charges that are induced in the Faraday cages, being given by $(-q/\hbar)\int dt (V_a-V_b)$. The final contribution comes from the interaction of the particle with the final vector potential $\mb{A}_f$ while its wave function propagates from the superposed positions $\mb{R}_a$ and $\mb{R}_b$ to the end position, being given by $(q/\hbar)[\int_{\mb{R}_a}^{\mb{R}_b} \mb{A}_f\cdot d\mb{x}]_\mathrm{end}$, with the integral performed from the path that goes from $\mb{R}_a$ to $\mb{R}_b$ passing through the end position in Fig. 3(a). By writing $\mb{A}_f=\mb{A}_i+\Delta\mb{A}$, we can write the total AB phase difference as
\begin{equation}\label{AB-gen1}
	\phi_{AB}^{(3)}= \frac{q}{\hbar}\int (V_b-V_a)dt+\frac{q}{\hbar}\oint\mb{A}_i\cdot d\mb{x}+\frac{q}{\hbar}\left[\int_{\mb{R}_a}^{\mb{R}_b}\Delta\mb{A}\cdot d\mb{x}\right]_\mathrm{end}
\end{equation}

The total AB phase difference given by Eq. (\ref{AB-gen1}) in general is nonzero. The second term on the right side of the above equation is zero, since no magnetic flux is enclosed by the paths. For very small Faraday cages, the first term on the right side should be negligible. But even for cages that are not so small, the first term on the right side may be negligible depending on the system symmetry. The total charge induced on the surface of each Faraday cage is zero, so in general there will be regions inside one cage with the same scalar potential in Lorenz gauge as regions inside the other cage (note that the scalar potential is not uniform inside the cages in this situation, since the electric field derived from it should cancel the electric field generated by the varying vector potential). The positions $\mb{R}_a$ and $\mb{R}_b$ can be chosen in these regions. On the other hand, the last term on the right side of Eq. (\ref{AB-gen1}) in general is nonzero. The fact that the vector potential changes during the particle propagation through the interferometer makes the circulation of the vector potential to be nonzero from the wavepackets point of view, generating this phase term. This proposed electrodynamic AB effect is thus distinct from the usual electric AB effect, since in the electric AB effect of Fig. 2 the origin of the AB phase (and of the system electric field) is a scalar potential, while in the electrodynamic AB effect of Fig. 3 the AB phase (and the system relevant electric field) is generated by a varying vector potential.

In Ref. \cite{singleton13}, the authors consider an AB scheme where the possible particle paths enclose an infinite solenoid whose electric current varies in time. But since there were no Faraday cages in this situation, the quantum particle suffers a path-dependent force due to the induced electric field, whose contribution cancels the AB phase generated by the current variation \cite{singleton13}. In our case, the absence of force due to the presence of the Faraday cages may result in a nonzero contribution from the electric current variation for the AB phase difference.


It is worth mentioning that, with multiparticle interference, there may be a nonzero AB phase even if no particle encloses a magnetic flux \cite{samulelson04,retzker06}. However, in these situations a combination of the different particles paths do enclose a magnetic flux \cite{samulelson04,retzker06}. So, we are dealing with a different situation in the electrodynamic AB effect proposed here. 

A nonzero AB phase difference in the scheme of Fig. 3(a) indicates that the AB phase is continuously acquired by the quantum particle during its propagation, as predicted in quantum electrodynamics treatments \cite{santos99,marletto20,saldanha21a,saldanha21b}.

A topological description for the AB phase difference of Eq. (\ref{AB-gen1}) in the scheme of Fig. 3(a) is also possible by considering the fields in spacetime, as shown in Fig. 3(b). The spacetime region corresponding to nonzero magnetic fields are represented in yellow and orange, with these two colors corresponding to different values for the field. The spacetime region corresponding to a nonzero  electric field is represented in green. We can see that the region with a null electromagnetic field is not simply-connected in spacetime. As before, we consider deformations of one of the possible particle trajectories into the other in spacetime, showing that the AB phase depends on the topology of the spacetime region where the electromagnetic fields are null. If it is not possible to deform one of the possible particle trajectories in spacetime into the other without crossing a region where the electromagnetic fields are nonzero, we may have a nonzero AB phase difference that depends on the flux of the electric and magnetic fields on the spacetime surface formed by this deformation. In the general case, the AB phase can be written as a combination of Eqs. (\ref{AB_B}) and (\ref{AB_E}) \cite{singleton13}:
\begin{equation}\label{AB_gen}
	\phi_{AB}=\frac{q}{\hbar}\left[\int_\mathcal{S}d\mb{a}(t)\cdot\mb{B}(\mb{r},t)-\int  \int_\mathcal{S} dt\,  d\mb{r}(t)\cdot\mb{E}(\mb{r},t)\right],
\end{equation}
with the integrals performed on the spacetime surface defined by the cited deformation of the interferometer possible trajectories. It is clear that the situations of Figs. 1 and 2, with the AB phase differences given by Eqs. (\ref{AB_B}) and (\ref{AB_E}), are particular situations that are accounted by the general form of Eq. (\ref{AB_gen}). Being written in terms of the electromagnetic fields, the AB phase of Eq. (\ref{AB_gen}) is clearly gauge-invariant.

In Eq. (\ref{AB-gen1}), by writing $V_b-V_a=-\int_{\mb{R}_a}^{\mb{R}_b}\mb{E}_V(\mb{r},t)\cdot d\mb{r}(t)$, where  $\mb{E}_V$ is the electric field generated by the scalar potential associated to the induced charges in the Faraday cages, and $\Delta\mb{A}=-\int dt \mb{E}_A$, where $\mb{E}_A$ is the electric field generated by the varying vector potential, we can write
\begin{equation}\label{AB_no}
	\phi_{AB}^{(3)}= -\frac{q}{\hbar}\left[\int dt \int_{\mb{R}_a}^{\mb{R}_b}\mb{E}(\mb{r},t)\cdot d\mb{r}\right]_\mathrm{end},
\end{equation}
$\mb{E}=\mb{E}_V+\mb{E}_A$ being the total electric field. In all regions outside the solenoid in Fig. 3(a), we have $\bs{\nabla}\times \mb{A}=0$ during all times, such that we can write $\mb{A}=\bs{\nabla}F$ with $F$ being a scalar function. Since we have $\mb{E}_V=-\bs{\nabla}V$ and $\mb{E}_A=-\partial/\partial t [\bs{\nabla}F]$, the spatial integral in Eq. (\ref{AB_no}) is the same for any two paths that do not enclose the solenoid. In this way, for the spacetime surface in blue in Fig. 3(b) constructed from the deformation of one of the possible particle trajectories into the other as described in Fig. 2(c), the AB phase computed by Eq. (\ref{AB_gen}) agrees with the one from Eq. (\ref{AB_no}) [note that the second integral of Eq. (\ref{AB_gen}) is performed in the spacetime surface formed by the superposition of the blue surface with the green solid in Fig. 3(b)]. This AB phase does not depend on the specific spacetime geometry of the possible particle trajectories. But to conclude that it is indeed topological, it is important to confirm that it does not depend on the specific spacetime surface that is formed with the cited trajectories deformation, as we do in the following.

As mentioned in the previous paragraph, the spatial integral in Eq. (\ref{AB_no}) is the same for any two paths that do not enclose the solenoid. But this is not true for 2 paths that, together, enclose the solenoid. Consider the spacetime surface in blue in Fig. 3(c), constructed from a different deformation of one of the possible particle trajectories into the other. For each fixed time, the spatial integral of the second term of Eq. (\ref{AB_gen}) from $\mb{R}_a$ to $\mb{R}_b$ in the spacetime surface of Fig. 3(c), when combined with the spatial integral from $\mb{R}_b$ to $\mb{R}_a$ in the spacetime surface of Fig. 3(b), form a closed curve that encloses the solenoid. This means that the contribution of the second term of Eq. (\ref{AB_gen}) for the AB phase difference differs for the spacetime surfaces of Figs. 3(c) and 3(b) by an amount
\begin{equation}\label{aux}
	\frac{q}{\hbar}\int_{t_i}^{t_f} dt \oint d\mb{r}\cdot\frac{\partial\mb{A}}{\partial t}= \frac{q}{\hbar}\left[\oint d\mb{r}\cdot\mb{A}\right]_{t_i}^{t_f}=\frac{q}{\hbar}(\Phi_f-\Phi_i),
\end{equation}
where we've used $\mb{E}_A=-\partial \mb{A}/\partial t$, $\bs{\nabla}\times\mb{A}=\mb{B}$, and Stokes' theorem, with $\Phi_i$ and $\Phi_f$ being the initial and final magnetic fluxes in the solenoid (at times $t_i$ and $t_f$). But note that in the spacetime surface of Fig. 3(c) the first term of Eq. (\ref{AB_gen}) also contributes for the AB phase, since we have two different magnetic fluxes crossing it. This contribution is given by $q(\Phi_i-\Phi_f)/\hbar$, canceling the difference of Eq. (\ref{aux}), such that we conclude that the AB phase difference  of Eq. (\ref{AB_gen}) is the same for the spacetime surfaces indicated in Figs. 3(b) and 3(c). It should be clear that this AB phase difference does not depend on the specific spacetime surface that is formed with the trajectories deformation in the scheme of Fig. 3(a). It depends only on the topology of the region with null electromagnetic field and on general properties of the possible particle trajectories in spacetime [such as if each possible trajectory crosses one of the ``holes'' in the green solid of Figs. 3(b) and 3(c)]. For this reason, the proposed electrodynamic AB effect of Fig. 3 is a topological effect. 

The extension of the presented calculations for situations where the potentials of the Faraday cages in Fig. 3(a) are controlled as in Fig. 2(a), or where the particle paths enclose a magnetic flux, is straightforward. In any case the general AB phase is given by Eq. (\ref{AB_gen}), written in terms of magnetic and electric fluxes in a surface in spacetime which has the particle possible trajectories as its boundary, in an explicitly gauge-invariant form.

Electronic Mach-Zehnder interferometers exist for electrons propagating in free space \cite{gronniger06,turner21} and in material media \cite{ji03}. If it is possible to include Faraday cages in the interferometer paths for one of these implementations, the solenoid current variation in the scheme of Fig. 3(a) could be time-correlated with the electrons incidence or with the electrons detections, in order to select the situations where the electrons only propagate in regions with negligible electromagnetic fields. It is worth mentioning that, since we have time-dependent fields in the scheme of Fig. 3(a), the Faraday cages in general cannot completely cancel the electric fields inside them. But for a sufficiently slow time variation of the solenoid current and for Faraday cages constructed with sufficiently good conductors, the resultant fields inside the cages should have a negligible influence in the particle dynamics, such that the AB phase difference in principle could be measured. But even if the fields are not completely canceled inside the Faraday cages, what happens is that extra effects would appear, and if they are properly taken into account, in principle the AB phase difference could still be inferred.

To summarize, we have presented the electrodynamic AB effect depicted in Fig. 3, where a nonzero AB phase difference may be present even if the interferometer paths enclose no magnetic flux and are subjected to a negligible scalar potential difference during the quantum charged particle propagation. The topological description of the effect was also presented, considering the possible particle trajectories in the interferometer and the electromagnetic field configuration in spacetime. It is clear that to understand the basic features of the AB effect, including the details of its topological nature, is essential for our fundamental understanding of the electromagnetic interactions in general. The experimental demonstration of the proposed AB effect of Fig. 3 would certainly be an important step in this direction.

The author greatly acknowledges Herman Batelaan and Ricardo Heras for very useful discussions. This work was supported by the Brazilian agencies CNPq (Conselho Nacional de Desenvolvimento Cient\'ifico e Tecnol\'ogico), CAPES (Coordenação de Aperfeiçoamento de Pessoal de Nível Superior), and FAPEMIG (Fundação de Amparo à Pesquisa do Estado de Minas Gerais).


%

\end{document}